\RequirePackage{fix-cm}
\documentclass[onecollarge]{svjour2}      
\smartqed  
\usepackage{amssymb,amsmath}
\usepackage{graphicx}
\usepackage{marvosym}
\newcommand{\envelope}{(\raisebox{-.5pt}{\scalebox{1.45}{\Letter}}\kern-1.7pt)}
 \journalname{Acta Mechanica}
\begin{document}

\title{The inverse problem of a mixed Li\'enard type nonlinear oscillator equation from symmetry perspective
}


\author{Ajey K. Tiwari \and S. N. Pandey \and V. K. Chandrasekar \and M. Senthilvelan \and M. Lakshmanan 
}


\institute{Ajey K. Tiwari \at
              Centre for Nonlinear Dynamics, School of Physics, Bharathidasan University, Tiruchirapalli - 620 024, India \\
              \email{ajey.nld@gmail.com}      \\     
            \emph{Present address:} 
            Astrophysics and Cosmology Research Unit, School of Mathematics, Statistics and Computer Science, University of KwaZulu-Natal, Private Bag X54001, Durban 4000, South Africa.  
           \and
           S. N. Pandey \at
              Department of Physics, Motilal Nehru National Institute of Technology, Allahabad - 211 004, India\\
              \email{snp@mnnit.ac.in}
              \and
              V. K. Chandrasekar  \at
              Centre for Nonlinear Science and Engineering, School of Electrical and Electronics Engineering, SASTRA University, Thanjavur - 613 401, India\\
              \email{chandru25nld@gmail.com}
              \and
              M. Senthilvelan \envelope \at
              Centre for Nonlinear Dynamics, School of Physics, Bharathidasan University, Tiruchirapalli - 620 024, India\\
              \email{velan@cnld.bdu.ac.in}
              \and
              M. Lakshmanan\at
              Centre for Nonlinear Dynamics, School of Physics, Bharathidasan University, Tiruchirapalli - 620 024, India\\
              \email{lakshman@cnld.bdu.ac.in}
}

\date{Received: date / Accepted: date}

\maketitle

\begin{abstract}
In this paper, we discuss the inverse problem for a mixed Li\'enard type nonlinear oscillator equation $\ddot{x}+f(x)\dot{x}^2+g(x)\dot{x}+h(x)=0$, where $f(x),\,g(x)$ and $h(x)$ are arbitrary functions of $x$. Very recently, we have reported the Lie point symmetries of this equation. By exploiting the interconnection between Jacobi last multiplier, Lie point symmetries and Prelle-Singer procedure we construct a time independent integral for the case exhibiting maximal symmetry from which we identify the associated conservative non-standard Lagrangian and Hamiltonian functions. The classical dynamics of the nonlinear oscillator is also discussed and certain special properties including isochronous oscillations are brought out.
\end{abstract}

\section{Introduction}
In general, the Lagrangian of an autonomous dynamical system is defined as the difference between kinetic and potential energies \cite{gold,gre,arn}. Systems which have Lagrangians of this form are termed as natural/standard Lagrangians. However, recent studies reveal that certain autonomous dynamical systems do admit other forms of Lagrangians in which no such clear identification of kinetic and potential energy terms can be made. In other words these new forms of Lagrangians do not have separate kinetic or potential energy terms. However, they provide the same equation of motion as that of Newton's equation of motion. These new forms of Lagrangians are termed as non-standard Lagrangians \cite{cari,vkc5}. The associated Hamiltonians are also called non-standard Hamiltonians where the canonically conjugate momenta assume quite non-standard forms. A well known example which admits non-standard Lagranigan/Hamiltonian is the damped harmonic oscillator \cite{glad11,glad} or the well known Mathews-Lakshmanan oscillator \cite{pmm}. Subsequent investigations have shown that several dynamical systems do admit non-standard Lagrangians \cite{mus,mus3,vkc}. Subsequently, attempts have also been made to quantize these non-standard Hamiltonians through various procedures \cite{chit,chit1,partha2,nuccinew,chi}.  

Recently, a few alternative approaches have been considered to derive standard as well as non-standard Lagrangians for nonlinear dynamical systems \cite{vkc5,nuccinew}. For a class of nonlinear oscillator equations the Jacobi last multiplier can often be utilized to derive the Lagrangian. The connection between Jacobi last multiplier and the Lagrangian was introduced by Jacobi \cite{partha,jac,jac2} and the potential applicability of this method has been brought out only recently by Nucci and her collaborators \cite{nucci,nucci2,nucci3}. To identify the multiplier one needs at least two distinct Lie point symmetries of the underlying equation of motion. The knowledge of Lie point symmetries then plays an important role in identifying the Lagrangian for the given nonlinear system \cite{nucci,nucci2,nucci3}. From a different perspective, we have shown that for one degree of freedom systems one can also identify time independent Hamiltonian from the time independent integral itself \cite{vkc6}. Identifying a time independent integral for a nonlinear dynamical system is itself a nontrivial problem. To overcome this problem, in this paper, we propose a method of deriving integral of motion when the Lie point symmetries are known. The underlying algorithm is as follows. From the known Lie point symmetries, we obtain the so called null forms and integrating factors associated with a given dynamical equation by utilizing the recently proposed interconnections between the null forms, integrating factors, Lagrange multiplier and Prelle-Singer procedure \cite{vkc3,vkc2,anna_third}. The null forms and integrating factors are then utilized to derive the integrals which in turn lead to the corresponding Lagrangian and Hamiltonian of the system. The procedure prescribed here is applicable for a class of systems of the form
\begin{eqnarray}
\ddot x=\phi(x,\dot x),\label{met1}
\end{eqnarray}
where over dot stands for differentiation with respect to $t$. The only assumption here is that Eq. (\ref{met1}) should admit atleast two parameter Lie point symmetries. To illustrate this idea, we consider an equation belonging to the mixed Li\'enard type equation \cite{akt}, that is 
\begin{eqnarray}
\ddot{x}+f(x)\dot{x}^2+g(x)\dot{x}+h(x)=0,\label{int111}
\end{eqnarray}
where $f(x),\,g(x)$ and $h(x)$ are all arbitrary functions of $x$. Specifically, in this paper, we consider the following integrable dynamical system admitting maximal number of eight symmetries, as deduced in Ref. \cite{akt},
\begin{align}
\ddot{x}+f(x)\dot{x}^2&+\left(k_1\Im(x)+k_2\right)\dot{x}+\frac{1}{F(x)}\left[\frac{k_1^2}{9}\Im(x)^3 \right. \nonumber\\
&\left. +\frac{k_1k_2}{3}\Im(x)^2+k_3\,\Im(x)+k_4\right]=0,\, F(x)=e^{\int{f(x)dx}},\,\Im(x)=\int{F(x)dx},\label{int4}
\end{align}
where ${k_i}^{'}s,\,i=1,2,3,4,$ are arbitrary parameters and $k_4=-\frac{2k_2^3}{9k_1}+\frac{k_2k_3}{k_1}$.  Obviously Eq. (\ref{int4}) belongs to the family (\ref{int111}). Recently, some special cases of the above equation have been analyzed in the literature in different perspectives, see for example Refs.\cite{new1,new2}. In this work, we consider a more generalized equation which encompasses all those special cases. Moreover establishing a conservative Hamiltonian description for a generalized equation like (\ref{int4}) is a tedious task and we have succeeded in getting that from the symmetries itself.

The plan of the paper is as follows. In Sec. 2, we briefly discuss the method of identifying suitable Lagrangian and Hamiltonian for the given second order ODE (\ref{int4}). The Lagrangian and Hamiltonian description of Eq. (\ref{int4}) are discussed in Sec. 3. In Sec. 4, we discuss the dynamical properties associated with (\ref{int4}) and prove its isochronicity. Finally, conclusion is given in Sec. 5.

\section{Methodology for the identification of Lagrangian}
In this section, we discuss the method for identifying Lagrangian for a second order ODE of the form (\ref{met1}). 
\subsection{First integral and Lagrangian}
In this section, we briefly recall the extended PS procedure \cite{vkc2}, which is one of the versatile tools to derive integrating factors and first integrals for the nonlinear second order ODEs, and introduce a method to obtain the corresponding Lagrangian and Hamiltonian.

Let us consider a second order ODE of the form $\ddot{x}=\phi(x,\dot{x})$, given by Eq. (\ref{met1}). We assume, that this ODE admits a first integral $I(t,x,\dot{x})=C$, where C is a constant, on the solutions. If $S(t,x,\dot{x})$ and $R(t,x,\dot{x})$ denote the null form and integrating factor for the second order ODE (\ref{met1}), then they can be determined from the relations (for more details one may refer \cite{vkc2})
\begin{subequations}
\label{ps}
\begin{align}
D[S]&=-\phi_{x}+S\phi_{\dot{x}}+S^{2},\label{ps8}\\
D[R]&=-R(S+\phi_{\dot{x}}), \label{ps9} \\
R_{x}&=R_{\dot{x}}S+RS_{\dot{x}}. \label{ps10}
\end{align}
\end{subequations}

Solving (\ref{ps}) one can obtain the expressions for $S$ and $R$. Once they are determined the associated integral can be obtained by evaluating the following expression \cite{vkc2}, that is
\begin{align}
I(t,x,\dot{x}) & =  G-\int \left( RS+\frac{d}{dx}G \right) dx-\int \Biggl\{R  +\frac{d}{d\dot{x}} \left[G- \int \left(RS+\frac{d}{dx}G\right)dx\right]\Biggl\}d\dot{x}.
  \label{ps11}
\end{align} 
where $G=\int R(\phi+\dot{x}S)dt$. Once we know the first integral we can obtain the Lagrangian and Hamiltonian corresponding to the equation with the help of the following method.

Suppose the given equation admits a time independent integral $I$ then one can associate it with a Hamiltonian \cite{vkc6} through the Legendre transform, namely
\begin{eqnarray}
I(x,\dot{x}) = H(x,p) = p \dot{x}-L(x,\dot{x}),\label{mlin13}
\end{eqnarray}
where $L(x,\dot{x})$ is the Lagrangian and $p$ is the canonically conjugate momentum. From (\ref{mlin13}), we can deduce
\begin{eqnarray}
\frac{\partial{I}}{\partial\dot{x}} = \frac{\partial{p}}{\partial\dot{x}} \dot{x}, \quad
\frac{\partial{I}}{\partial{x}} = \frac{\partial{p}}{\partial{x}}\dot{x}-\frac{\partial{L}}{\partial{x}}.\label{mlin13a}
\end{eqnarray}
Identifying without loss of generality $p$ and $L$ from (\ref{mlin13a}), we find
\begin{align}
p &=\int \frac{I_{\dot{x}}}{\dot{x}} d\dot{x},\nonumber\\
L &= \int (p_x\dot{x}-I_{x})dx + \int [p-\frac{d}{d\dot{x}}\int (p_x\dot{x}-I_{x})dx]d\dot{x}.\label{mlin13b}
\end{align}
Using expressions (\ref{mlin13})-(\ref{mlin13b}) we can obtain the expressions for the Lagrangian, the canonically conjugate momentum and the associated Hamiltonian. 

Hence, from the above study it is clear that once we know the expressions for the null form and integrating factor we can obtain the first integral that will lead to the Lagrangian/Hamiltonian of the system. However, in some cases, it is difficult to solve the determining equations for the null forms and integrating factor. To overcome this difficulty, we analyze the recently established interconnections between the Lie point symmetries, null form, integrating factor and Jacobi multiplier that will directly give the expressions for the null form and integrating factor.  

\subsection{Exploring the interconnections}
\subsection*{(i) Lie symmetries and null forms}
Let us consider the Lie point symmetry generator of (\ref{met1}) be $V=\xi\partial_t+\eta\partial_x$, where $\xi(t,x)$ and $\eta(t,x)$ are the symmetry functions. The second prolonged vector field of $V$ is given by $V^{(2)}(=\xi\partial_t+\eta\partial_x+\eta^{(1)}\partial_{\dot{x}}+\eta^{(2)}\partial_{\ddot{x}}),$ where $\eta^{(1)}$ and $\eta^{(2)}$ are the first and second prolongations of the vector field $V$ \cite{olver}. Here, $\eta^{(1)}=\dot{\eta}-\dot{x}\dot{\xi}$ and $\eta^{(2)}=\ddot{\eta}-\ddot{x}\dot{\xi}$. The infinitesimal symmetries $\xi$ and $\eta$ can be determined by imposing the invariance of the equation of motion (\ref{met1}) under the infinitesimal transformation $T=t+\epsilon \xi(t,x)$ and $X=x+\epsilon \eta(t,x)$ or equivalently 
\begin{eqnarray}
V^{(2)}[\ddot{x}-\phi(t,x,\dot{x})]\mid_{\ddot{x}-\phi(t,x,\dot{x})=0}=0.\label{num1}
\end{eqnarray}
Alternatively, the above invariance condition can be formulated in terms of the characteristic of $V$, that is $Q(=\eta-\dot{x}\xi)$ as
\begin{equation}  
D^2[Q]=\phi_{\dot{x}} D[Q]+\phi_x Q.\label{met16}
\end{equation}
Solving Eq. (\ref{met16}) we can get $Q$ from which we can extract Lie point symmetries $\xi$ and $\eta$ with appropriate restrictions on $Q$.

Now, introducing the following transformation to the null function,
\begin{equation}
S=-D[X]/X\label{sx},
\end{equation} 
Eq.(\ref{ps8}) becomes a linear equation in the new variable $X$, that is
\begin{equation}  
D^2[X]  = \phi_{\dot{x}} D[X]+\phi_x X,\label{met14}
\end{equation}
where $D$ is the total differential operator. Comparing Eqs.(\ref{met14}) and (\ref{met16}) we find that
\begin{equation}
X=Q.\label{xq}
\end{equation} 
Interestingly, the $S$-determining equation (\ref{met14}) is exactly the same as that of the determining equation for the Lie point symmetries (\ref{met16}). Since $S=-\frac{D[X]} {X}$, the null form $S$ can also be determined once $\xi$ and $\eta$ are known (for more details one may refer \cite{vkc3}).

\subsection*{(ii) Integrating factor and Jacobi last multiplier}
To get the interconnection between the Jacobi last multiplier and integrating factor we rewrite Eq. (\ref{met1}) into an equivalent system of two first-order ODEs $\dot{x}_i=w_i(x_1,x_2)$, $i=1,2$. Then its Jacobi last multiplier $M$ is obtained by solving the following differential equation \cite{nucci3},
\begin{equation}  
\frac{\partial M}{\partial t}+\sum_{i=1}^2\frac{\partial (Mw_i)}{\partial x_i}=0\label{met18}.
\end{equation}
The above equation can be rewritten as
\begin{equation}  
D[\log M]+\phi_{\dot{x}}=0.\label{met20}
\end{equation}
Introducing now the following transformation for the integrating factor
\begin{equation}
R=X/F,\label{rxf}
\end{equation}
where $F(t,x,\dot{x})$ is a function to be determined, in Eq. (\ref{ps9}) we can rewrite the latter as
\begin{equation}  
D[F] = \phi_{\dot{x}}F.
\label{met15}
\end{equation}
Now comparing Eq. (\ref{met15}) with (\ref{met20}) we find that $F=M^{-1}$. Thus from the knowledge of the multiplier $M=\big(\frac{1} {\Delta}\big)$ we can also fix the explicit form of $F$ which appears in the denominator of integrating factor $R$ (vide Eq.(\ref{rxf})) as $F=\Delta$ or vice versa. Here, the multiplier $M$ of a second order ODE can be obtained as \cite{vkc3}
\begin{eqnarray}
\frac{1}{M}=\Delta=\left|
\begin{matrix}
    1 & \dot{x} & \ddot{x} \\
    \xi_1 & \eta_1 & \eta_1^{(1)} \\
    \xi_2 & \eta_2 & \eta_2^{(1)}
\end{matrix} \right|,\label{int2.1}
\end{eqnarray}
where, $(\xi_1,\eta_1)$ and $(\xi_2,\eta_2)$ are two sets of Lie point symmetries of the ODE (\ref{met1}) and $\eta_1^{(1)}$ and $\eta_2^{(1)}$ are their corresponding first prolongations, that is $\eta_i^{(1)}=\dot{\eta_i}-\dot{x}\dot{\xi_i},\,i=1,2$. 

We also note here that once the Jacobi last multiplier is found the Lagrangian can be constructed directly from the relation
\begin{eqnarray}
\frac{\partial^2L}{\partial \dot{x}^2}=\frac{1}{\Delta}=M.\label{int2.2}
\end{eqnarray}
Integrating the multiplier two times with respect to $\dot{x}$ we can obtain a Lagrangian for the given equation. However, as we see below, the obtained Lagrangians for some problems turn out to be too messy and it is difficult to use the underlying expression.

\section{Lagrangian and Hamiltonian description of Eq. (\ref{int4})}
In this section, we illustrate the effectiveness of the procedure discussed in the previous section by identifying the non-standard Lagrangian and Hamiltonian for Eq. (\ref{int4}).

\subsection{Time independent integral}
Eq. (\ref{int4}) is invariant under the following two Lie point symmetries \cite{akt}, namely
\begin{subequations}
\label{ggg}
\begin{align}
\xi_1&=1,\qquad \eta_1=0.\label{sym4.1}\\
\xi_2&=-\frac{k_1\Im(x)+k_2}{3k_3-k_2^2},~~
\eta_2=\frac{k_1\Im(x)+k_2}{k_1F(x)}+\frac{\left(k_1\Im(x)+k_2\right)^3}{3k_1F(x)\left(3k_3-k_2^2\right)},\label{sym4}
\end{align}
\end{subequations}
where $F(x)=e^{\int{f(x)dx}}$ and $\Im(x)=\int{F(x)dx}$. One may also consider the other Lie point symmetries of Eq. (\ref{int4}) to derive the Lagrangian. Since we are interested to explore the time independent integrals we consider only these two symmetries. To determine the null forms we consider the Lie point symmetries (\ref{sym4.1}) from which we can fix $Q=\eta-\dot x \xi=-\dot{x}$. Substituting the latter expression and its derivative in (\ref{sx}), we readily obtain
\begin{align}
S=\frac{f\dot{x}^2+(k_1\Im+k_2)\dot{x}+\frac{k_1^2}{9}\frac{\Im^3}{F}+\frac{k_1k_2}{3}\frac{\Im^2}{F}+\frac{k_3\Im}{F}+\frac{k_4}{F}}{\dot{x}},\label{int02.21}
\end{align}
where $k_4=\frac{9k_2k_3-2k_2^3}{9k_1F}$. From the knowledge of $S,\,X$ and the Lie point symmetries (\ref{sym4.1}) we can find the integrating factor $R$ to be of the form (see Eq. (\ref{rxf}))
\begin{eqnarray}
R=\frac{54k_1^2F^2(k_2-3k_3)\dot{x}}{P_1P_2},\label{int1012.21}
\end{eqnarray}
where $P_1$ and $P_2$ are given as
\begin{subequations}
\label{sym7-8}
\begin{eqnarray}
P_1&=&\left((3k_1F(x)\dot{x}+\left(k_1\Im(x)+k_2\right)^2\right)^2-3\left(k_2^2-3k_3\right)\left(k_1\Im(x)+k_2\right)^2,\label{sym7}\\
P_2&=&(k_1\Im(x)+k_2)^2-3(k_2^2-3k_3)+3k_1F(x)\dot{x}.\label{sym8}
\end{eqnarray}
\end{subequations}
The integral of motion of (\ref{int4}) can now be obtained by substituting (\ref{int02.21}) and (\ref{int1012.21}) in (\ref{ps11}) and evaluating the integrals. Doing so, we find a time independent integral
\begin{align}
I_1=\frac{P_1}{P_2^2},\label{int02.22}
\end{align}
where $P_1$ and $P_2$ are given in (\ref{sym7-8}). 

\subsection{Conservative Lagrangian/Hamiltonian}
To derive the time independent Lagrangian and Hamiltonian from the first integral (\ref{int02.22}), we first determine the canonical momentum from the first integral (\ref{int02.22}). Substituting the latter in the first of the relations in (\ref{mlin13b}) and integrating the resultant expression, we find
\begin{align} 
p&=-\frac{27\lambda^3}{2k_1}\left(\frac{F(x)}{\left(k_1F(x)\dot{x}+\frac{k_1^2}{3}\left(\frac{k_2}{k_1}+\Im(x)\right)^2+3\lambda\right)^2}\right)+\frac{3\lambda}{2k_1}F(x),\label{ch5.a6}
\end{align}
where $\lambda=k_3-\frac{k_2^2}{3}$.  Once $p$ is known the Lagrangian and Hamiltonian can be determined by evaluating the expression (\ref{mlin13b}). Our result shows ($\lambda>0$)
\begin{align} 
L&=\frac{27\lambda^3}{2k_1^2}\left(\frac{1}{k_1F(x)\dot{x}+\frac{k_1^2}{3}\left(\frac{k_2}{k_1}+\Im(x)\right)^2+3\lambda}\right)+\frac{3\lambda}{2k_1}F(x)\dot{x}-\frac{9\lambda^2}{2k_1^2}\label{ch5.a5}
\end{align}
and 
\begin{align} 
H=\frac{9\lambda^2}{2}\left[\frac{\left(F(x)\dot{x}+\frac{k}{3}\left(\frac{k_2}{k_1}+\Im(x)\right)^2\right)^2+\lambda \left(\frac{k_2}{k_1}+\Im(x)\right)^2}{\left(k_1F(x)\dot{x}+\frac{k_1^2}{3}\left(\frac{k_2}{k_1}+\Im(x)\right)^2+3\lambda\right)^2}\right],\label{ch5.a7.1}
\end{align}
or
\begin{align} 
H=\frac{9\lambda^2}{2k_1^2}\left[2-\frac{2k_1p}{3\lambda F}-2\,\left(1-\frac{2k_1p}{3\lambda F}\right)^{\frac{1}{2}}+\frac{k_1^2}{9\lambda}\left(\frac{k_2}{k_1}+\Im(x)\right)^2\right.
\left. \left(1-\frac{2k_1p}{3\lambda F}\right)\right].\label{ch5.a7.2}
\end{align}
One may observe that both the Lagrangian and Hamiltonian are of non-standard forms. One can check that the Euler-Lagrange equation of motion derived from (\ref{ch5.a5}) and the Hamiltonian equation of motions derived from (\ref{ch5.a7.1}) coincide with that of Eq. (\ref{int4}). For the case $\lambda<0$ Eq. (\ref{int4}) admits the same Lagrangian and Hamiltonian descriptions given by Eqs. (\ref{ch5.a5}) and (\ref{ch5.a7.1}) or (\ref{ch5.a7.2}), respectively, provided that one has to replace $\lambda$ with $-\mid \lambda \mid$ in the respective places in Eqs. (\ref{ch5.a5}) and (\ref{ch5.a7.1}) or (\ref{ch5.a7.2}).

\subsection{Alternative Lagrangian}
As we pointed out in the previous section, the Lagrangian can also be determined by determining Jacobi last multiplier $M$ (see Eq. (\ref{int2.2})). Substituting the Lie point symmetries (\ref{ggg}) in (\ref{int2.1}) and evaluating the determinant, we find

\begin{eqnarray}
M&=&\frac{1}{27k_1^2(k_2^2-3k_3)F(x)^2}\Bigl[\left(k_1\Im(x)+k_2\right)\left(\left(k_1\Im(x)+k_2\right)^2-3(k_2^2-3k_3)\right) \Bigl\{\left(k_1\Im(x)+k_2\right)\nonumber \\
&&\left(\left(k_1\Im(x)+k_2\right)^2-3(k_2^2-3k_3)\right)+9k_1F(x)\dot{x}\left(k_1\Im(x)+k_2+f(x)\dot{x}\right)\Bigr\}\nonumber \\
&&+9k_1F(x)\dot{x}^2\Bigl\{3k_1F(x)\left(\left(k_1\Im(x)+k_2\right)^2-(k_2^2-3k_3)+k_1F(x)\dot{x}\right)\nonumber \\ &&
-f(x)\left(k_1\Im(x)+k_2\right)\left(\left(k_1\Im(x)+k_2\right)^2-3(k_2^2-3k_3)\right)\Bigr\}\Bigr],\label{sym5}
\end{eqnarray}
where $F=e^{\int{f(x)dx}}$ and $\Im(x)=\int{F(x)dx}$. Substituting the multiplier $M$ in Eq. (\ref{int2.2}) and integrating it twice with respect to $\dot{x}$, we obtain
\begin{align}
L&=\frac{1}{2P_3}  \left[P_2\ln{\frac{P_1}{P_2^2}}-\frac{6\sqrt{3}k_1F(x)\sqrt{k_2^2-3k_3}}{k_1\Im(x)+k_2}\right.\left. \times\tanh^{-1}\left\{\frac{(k_1\Im(x)+k_2)^2+3k_1F(x)\dot{x}}{\sqrt{3}(k_1\Im(x)+k_2)\sqrt{k_2^2-k_3}}\right\}\right],\label{sym6}
\end{align}
where $P_3=(k_1\Im(x)+k_2)^2-3(k_2^2-3k_3)$. One can check that Eq. (\ref{sym6}) is another non-standard Lagrangian for Eq. (\ref{int4}). 

\section{Hamiltonian Dynamics of Eq. (\ref{int4})}
For the present problem, it is very difficult to obtain the second integral either through PS procedure or by recalling the interconnections so as to obtain the explicit solutions and study the underlying dynamics. It is also clear from the form of Eq. (\ref{int4}) that it cannot be integrated directly. So we have to follow another procedure to obtain the solution and explore its underlying dynamics. Since the equation has a conservative Hamiltonian description the best way to do this is to identify a canonical transformation which can transform the complicated Hamiltonian into a simple Hamiltonian from which the solutions can be obtained. After a careful analysis we identified a canonical transformation that transforms the Hamiltonian (\ref{ch5.a7.1}) into a harmonic oscillator type Hamiltonian.

Introducing the canonical transformation
\begin{align}
\int{e^{\int{f(x)dx}}dx}&=-\frac{k_2}{k_1}+\frac{U}{1-\frac{k_1}{3\lambda}P},\nonumber\\
 p&=P e^{-\int{f(x)dx}}\left(1-\frac{k_1}{6\lambda}P\right),\label{ch5.a8}
\end{align}
where $P=\dot{U}$, the Hamiltonian (\ref{ch5.a7.1}) can be transformed to the linear harmonic oscillator Hamiltonian $H=\frac{1}{2}\left(P^2+\lambda U^2\right)$. It is clear that the parameter $\lambda$ fixes the ultimate dynamics of the dynamical system (\ref{int4}). We consider all the three cases, namely $(i)\,\lambda>0,\,(ii)\,\lambda<0$ and $(iii)\,\lambda=0$ and bring out the underlying dynamics of Eq. (\ref{int4}). 
\subsection{ Case $(i)\,{\lambda>0}$}
In the first case $\lambda>0$, we have $\ddot U+\lambda U=0$ which upon integration gives $U=A\sin(\omega t+\delta)$, where $A$ and $\delta$ are arbitrary constants. From the solution of the latter we arrive at the following sinusoidal solution for Eq. (\ref{int4}) of the form
\begin{eqnarray}
\int{e^{\int{f(x)dx}}dx}=-\frac{k_2}{k_1}+\frac{A\sin{(\omega t+\delta)}}{1-(\frac{k_1}{3\omega})A\cos{(\omega t+\delta)}},\label{int40}
\end{eqnarray}
where $ 0\leq A <\frac{3\omega}{k_1},\,\,\omega=\sqrt{\lambda}$ and $A=\frac{\sqrt{2H}}{\omega}$ and $\delta$ is an arbitrary constant. It is to be noted that the above general solution admits harmonic periodic solutions and that the frequency of the periodic oscillations is completely independent of the amplitude or initial condition. Consequently, the underlying system turns out to be an isochronous one. This surprising result is not a common feature for nonlinear oscillators. 

\subsection{ Case (ii)\,${\lambda<0}$}
For the case $\lambda<0$, Eq. (\ref{int4}) admits the same Lagrangian and Hamiltonian descriptions given by Eqs. (\ref{ch5.a5}) and (\ref{ch5.a7.1}) or (\ref{ch5.a7.2}), respectively. The only difference here is that one has to replace $\lambda$ with $-\mid \lambda \mid$ in the respective places in Eqs. Eqs. (\ref{ch5.a5}) and (\ref{ch5.a7.1}) or (\ref{ch5.a7.2}). Hence, the solution for Eq. (\ref{int4}) for the case $\lambda<0$ can be written by solving the underlying equation of motion which in turn provides us the solution of (\ref{int4}) in the form
\begin{align}
\int{e^{\int{f(x)dx}}dx}=-\frac{k_2}{k_1}+\frac{3\alpha\left(e^{\alpha\left(t-\delta\right)}-2Ee^{-\alpha\left(t-\delta\right)}\right)}{3\alpha+k_1\left(e^{\alpha\left(t-\delta\right)}+2Ee^{-\alpha\left(t-\delta\right)}\right)},\label{int41}
\end{align}
where $H=E,\,\alpha=\sqrt{\lvert{\lambda}\rvert}$ and $\delta $ is a constant. The above solution admits decaying type or aperiodic solutions only. The dissipative nature of the system is clear from (\ref{int41}).

\subsection{ Case $(iii)\,{\lambda=0}$}
Finally, we consider the case $\lambda=0$. In this case the Hamiltonian can be written as 
\begin{align}
\hat{H}&=\frac{k_1}{3}\,p\,\left(\frac{k_2}{k_1}+\int{e^{\int{f(x)dx}}dx}\right)^2e^{-\int{f(x)dx}}-2\sqrt{-\frac{1}{k_1}\,p\,e^{-\int{f(x)dx}}},\label{int42}
\end{align}
where 
\begin{eqnarray}
p=-\frac{k_1e^{\int{f(x)dx}}}{\left(k_1e^{\int{f(x)dx}}\dot{x}+\frac{k_1^2}{3}\left(\frac{k_2}{k_1}+\int{e^{\int{f(x)dx}}dx}\right)^2\right)^2}.\label{int43}
\end{eqnarray}
For this case we identify the canonical transformation as
\begin{align}
U&=\sqrt{\frac{6}{k_1}\,p\,e^{-\int{f(x)dx}}},\nonumber\\
P&=\frac{1}{3}\sqrt{\frac{6}{k_1}\,p\,e^{-\int{f(x)dx}}}\left(\frac{k_2}{k_1}+\int{e^{\int{f(x)dx}}dx}\right).\label{cat1a}
\end{align}
The canonical transformation (\ref{cat1}) transforms (\ref{int42}) to a freely falling particle Hamiltonian $H=\frac{P^2}{2}-\sqrt{\frac{2}{3}}U$. Consequently the general solution to (\ref{int4}) for $\lambda=0$ can be given as
\begin{eqnarray}
\int{e^{\int{f(x)dx}}dx}=-\frac{k_2}{k_1}+\frac{3}{k_1}\left(\frac{2\,\left(t+I_1\right)}{\left(t+I_1\right)^2-3E}\right),\qquad H=E,
\end{eqnarray}
where $I_1$ is a constant.  

\section{Examples:}
We have shown above that system (\ref{int4}) exhibits time independent non-standard Lagrangian and Hamiltonian structures and also proved its isochronous nature for arbitrary form of the function $f(x)$. The implications can be easily appreciated by considering specific forms for the function $f(x)$. Here, in this section, we obtain time independent Lagrangian and conservative Hamiltonian for two specific choices of $f(x)$. 

\subsection{$f(x)=\frac{2\alpha x}{1+\alpha x^2}$}
For the choice $f(x)=\frac{2\alpha x}{1+\alpha x^2}$, Eq. (\ref{int4}) takes the form
\begin{equation}
\ddot{x}+\frac{2\alpha x}{1+\alpha x^2}\dot{x}^2+\bigg(k_1 \bigg(x+\frac{\alpha x^3}{3}\bigg)+k_2\bigg)\dot{x}+\frac{1}{1+\alpha x^2}\bigg[\frac{k_1^2}{9}\bigg(x+\frac{\alpha x^3}{3}\bigg)^3+\frac{k_1 k_2}{3}\bigg(x+\frac{\alpha x^3}{3}\bigg)^2+k_3\bigg(x+\frac{\alpha x^3}{3}\bigg)+k_4\bigg]=0,\label{exm1}
\end{equation}
where we choose $k_4=-\frac{2k_2^3}{9k_1}+\frac{k_2k_3}{k_1}$. The conventional prediction about Eq. (\ref{exm1}) is that, it is of dissipative nature. However, using the results given in Sec. 2, we can prove that the system (\ref{exm1}) is of conservative nature and displays the explicit forms of the time independent Lagrangian and Hamiltonian. The underlying expressions are given by
\begin{subequations}
\begin{align} 
L&=\frac{27\lambda^3}{2k_1^2}\left(\frac{1}{k_1(1+\alpha x^2)\dot{x}+\frac{k_1^2}{3}\left(\frac{k_2}{k_1}+x+\frac{\alpha x^3}{3}\right)^2+3\lambda}\right)+\frac{3\lambda}{2k_1}(1+\alpha x^2)\dot{x}-\frac{9\lambda^2}{2k_1^2}\label{exm5}\\
H&=\frac{9\lambda^2}{2}\left[\frac{\left((1+\alpha x^2)\dot{x}+\frac{k}{3}\left(\frac{k_2}{k_1}+x+\frac{\alpha x^3}{3}\right)^2\right)^2+\lambda \left(\frac{k_2}{k_1}+x+\frac{\alpha x^3}{3}\right)^2}{\left(k_1(1+\alpha x^2)\dot{x}+\frac{k_1^2}{3}\left(\frac{k_2}{k_1}+x+\frac{\alpha x^3}{3}\right)^2+3\lambda\right)^2}\right].\label{exm7.2}
\end{align}
\end{subequations}
It is clear from the forms of $L$ and $H$ that both are of non-standard type as the $x$ and $p$ components are mixed up such that is one can not make a clear distinction between the kinetic and potential energy terms. Due to such form its seems difficult to predict the dynamics of the system completely. However, using the canonical transformation
\begin{subequations}
\label{ch5.aa8}
\begin{align}
x+\frac{\alpha x^3}{3}&=-\frac{k_2}{k_1}+\frac{U}{1-\frac{k_1}{3\lambda}P},\\
 p(1+\alpha x^2)&=P \left(1-\frac{k_1}{6\lambda}P\right),
\end{align}
\end{subequations}
we can transform the above Hamiltonian to that of the linear harmonic oscillator equation from which the solution for Eq. (\ref{exm1}) can be written for three different choices as follows.
\subsection*{Case $(i)$: $k_3>0$}
For $k_3>0$ the form of the solution (\ref{int40}) for the equation (\ref{exm1}) turns out to be 
\begin{eqnarray}
x+\frac{\alpha x^3}{3}=-\frac{k_2}{k_1}+\frac{A\sin{(\omega t+\delta)}}{1-(\frac{k_1}{3\omega})A\cos{(\omega t+\delta)}},\label{int401}
\end{eqnarray}
Solving the above equation, we get three roots. In fact, we have only one real root and it is given by
\begin{eqnarray}
x(t)=\frac{2^{\frac{1}{3}}}{(-3\beta \alpha^2+(\sqrt{4\alpha^3+9\beta^2\alpha^4}))^{\frac{1}{3}}}-\frac{(-3\beta \alpha^2+(\sqrt{4\alpha^3+9\beta^2\alpha^4}))^{\frac{1}{3}}  }{2^{\frac{1}{3}}\alpha},
\end{eqnarray}
where $\beta=-\frac{k_2}{k_1}+\frac{A\sin{(\omega t+\delta)}}{1-(\frac{k_1}{3\omega})A\cos{(\omega t+\delta)}}$, $0\leq A <\frac{3\omega}{k_1}$ and $\omega=\sqrt{k_3}$. It is clear from the form of the solution that the dynamics of Eq. (\ref{exm1}) for $k_3>0$ is periodic and the frequency of the oscillations is independent of amplitude, that is the system is isochronous. 

\subsection*{Case $(ii)$: $k_3<0$}
For $k_3<0$, the solution for  Eq. (\ref{exm1}) can be written as
\begin{align}
x(t)=\frac{2^{\frac{1}{3}}}{(-3\gamma \alpha^2+(\sqrt{4\alpha^3+9\gamma^2\alpha^4}))^{\frac{1}{3}}}-\frac{(-3\gamma \alpha^2+(\sqrt{4\alpha^3+9\gamma^2\alpha^4}))^{\frac{1}{3}}  }{2^{\frac{1}{3}}\alpha},\label{int41a}
\end{align}
where $\gamma=-\frac{k_2}{k_1}+\frac{3\alpha\left(e^{\alpha\left(t-\delta\right)}-2Ee^{-\alpha\left(t-\delta\right)}\right)}{3\alpha+k_1\left(e^{\alpha\left(t-\delta\right)}+2Ee^{-\alpha\left(t-\delta\right)}\right)}$, $H=E,\,\alpha=\sqrt{\lvert{\lambda}\rvert}$ and $\delta $ is a constant. Here also we have three roots for Eq.(\ref{exm1}). We have considered only the real root. The form of the solution suggests decaying type or aperiodic solution for Eq. (\ref{exm1}). 

\subsection*{Case $(ii)$: $k_3=0$}
For the choice $k_3=0$, the Hamiltonian can be written as
\begin{align}
\hat{H}&=\frac{k_1}{3}\,\left(\frac{k_2}{k_1}+x+\frac{\alpha x^3}{3}\right)^2\frac{p}{1+\alpha x^2}-2\sqrt{-\frac{1}{k_1}\,\frac{p}{1+\alpha x^2}},\label{int42a}
\end{align}
where 
\begin{eqnarray}
p=-\frac{k_1(1+\alpha x^2)}{\left(k_1(1+\alpha x^2)\dot{x}+\frac{k_1^2}{3}\left(\frac{k_2}{k_1}+x+\frac{\alpha x^3}{3}\right)^2\right)^2}.\label{int43a}
\end{eqnarray}
For this case we identify the canonical transformation as
\begin{subequations}
\label{cat1}
\begin{align}
U&=\sqrt{\frac{6}{k_1}\,\frac{p}{1+\alpha x^2}},\\
P&=\frac{1}{3}\sqrt{\frac{6}{k_1}\,\frac{p}{1+\alpha x^2}}\left(\frac{k_2}{k_1}+x+\frac{\alpha x^3}{3}\right).
\end{align}
\end{subequations}
The canonical transformation (\ref{cat1}) transforms (\ref{int42}) to a freely falling particle Hamiltonian $H=\frac{P^2}{2}-\sqrt{\frac{2}{3}}U$. Consequently, the general solution to (\ref{exm1}) for $k_3=0$ can be given as
\begin{eqnarray}
x=\frac{2^{\frac{1}{3}}}{(-3a_1(t) \alpha^2+(\sqrt{4\alpha^3+9a_1(t)^2\alpha^4}))^{\frac{1}{3}}}-\frac{(-3a_1(t) \alpha^2+(\sqrt{4\alpha^3+9a_1(t)^2\alpha^4}))^{\frac{1}{3}}  }{2^{\frac{1}{3}}\alpha},\quad H=E,\label{aaa112}
\end{eqnarray}
where $a_1(t)=-\frac{k_2}{k_1}+\frac{3}{k_1}\left(\frac{2\,\left(t+I_1\right)}{\left(t+I_1\right)^2-3E}\right)$ and $I_1$ is a constant. The front-like form of the solution is clear from Eq. (\ref{aaa112}).

Hence, we can conclude that Eq. (\ref{exm1}) exhibits both conservative and front-like like behaviours depending on the value of the parameter $k_3$. For $k_3>0$ Eq. (\ref{exm1}) exhibits isochronous nature; however for $k_3\leq 0$ we get front-like nature.

\subsection{$f(x)=0$}
For this choice, Eq. (\ref{int4}) becomes modified Emden equation, that is
\begin{align}
\ddot{x}+k_1x\dot{x}+\frac{k_1^2}{9}x^3+k_3x=0,\label{exma1}
\end{align}
where we have chosen $k_2=0$. The above equation is well known in the literature and has been studied by many authors in different perspectives \cite{cari,vkc5,chit,partha2,subha}. Using the results given in Sec. 3, we can write expressions for time independent Lagrangian and Hamiltonian as
\begin{subequations}
\begin{align} 
L&=\frac{27k_3^3}{2k_1^2}\left(\frac{1}{k_1\dot{x}+\frac{k_1^2}{3}x^2+3k_3}\right)
+\frac{3k_3}{2k_1}\dot{x}-\frac{9k_3^2}{2k_1^2},\label{exma5}\\
H&=\frac{9k_3^2}{2k_1^2}\left[2-\frac{2k_1p}{3k_3 }-2\,\left(1-\frac{2k_1p}{3k_3}\right)^{\frac{1}{2}}
 +\frac{k_1^2}{9k_3}x^2 \left(1-\frac{2k_1p}{3k_3x}\right)\right],\\
p&=-\frac{27\lambda^3}{2k_1}\left(\frac{1}{\left(k_1\dot{x}+\frac{k_1^2}{3}\left(\frac{k_2}{k_1}+x\right)^2+3\lambda\right)^2}\right)
+\frac{3\lambda}{2k_1}.
\label{exma7.2}
\end{align}
\end{subequations}
Using the canonical transformation
\begin{subequations}
\label{ach5.aa8}
\begin{align}
x&=\frac{U}{1-\frac{k_1}{3k_3}P},\\
p&=P\left(1-\frac{k_1}{6k_3}P\right),
\end{align}
\end{subequations}
we can transform the above Hamiltonian to linear harmonic oscillator equation from which the solution for Eq. (\ref{exma1}) can be written for three different choices as follows.
\subsection*{Case $(i)$: $k_3>0$}
For $k_3>0$ the solution turns out to be of the form \cite{vkc5}
\begin{eqnarray}
x(t)=\frac{A\sin{(\omega t+\delta)}}{1-(\frac{k_1}{3\omega})A\cos{(\omega t+\delta)}},
\end{eqnarray}
where $0\leq A <\frac{3\omega}{k_1}$ and $\omega=\sqrt{k_3}$. It is clear from the form of the solution that the dynamics of Eq. (\ref{exma1}) for $k_3>0$ is periodic and the frequency of the oscillations is independent of amplitude, that is the system is isochronous. 

\subsection*{Case $(ii)$: $k_3<0$}
For $k_3>0$, the solution for  Eq. (\ref{exma1}) can be written as \cite{vkc5}
\begin{align}
x(t)=\frac{3\alpha\left(e^{\alpha\left(t-\delta\right)}-2Ee^{-\alpha\left(t-\delta\right)}\right)}{3\alpha+k_1\left(e^{\alpha\left(t-\delta\right)}+2Ee^{-\alpha\left(t-\delta\right)}\right)},\label{aint41a}
\end{align}
where $H=E,\,\alpha=\sqrt{\lvert{\lambda}\rvert}$ and $\delta $ is a constant. The form of the solution suggests decaying/front-like type solutions for Eq. (\ref{exma1}). 

\subsection*{Case $(ii)$: $k_3=0$}
For the choice $k_3=0$, the Hamiltonian can be written as
\begin{align}
\hat{H}&=\frac{k_1}{3}{px^2}-2\sqrt{-\frac{p}{k_1}},\label{aint42a}
\end{align}
where 
\begin{eqnarray}
p=-\frac{k_1}{\left(k_1\dot{x}+\frac{k_1^2}{3}x^2\right)^2}.\label{aint43a}
\end{eqnarray}
For this case we identify the canonical transformation as
\begin{subequations}
\label{acat1}
\begin{align}
U&=\sqrt{\frac{6}{k_1}\,p},\\
P&=\frac{1}{3}\sqrt{\frac{6}{k_1}\,p}\,\,x.
\end{align}
\end{subequations}
The canonical transformation (\ref{acat1}) transforms (\ref{aint42a}) to a freely falling particle Hamiltonian $H=\frac{P^2}{2}-\sqrt{\frac{2}{3}}U$. Consequently, the general solution to (\ref{exma1}) for $k_3=0$ can be given as
\begin{eqnarray}
x(t)=\frac{3}{k_1}\left(\frac{2\,\left(t+I_1\right)}{\left(t+I_1\right)^2-3E}\right),\qquad H=E,\label{aaaa112}
\end{eqnarray}
where $I_1$ is a constant. The dissipative form of the solution is clear from Eq. (\ref{aaaa112}).

Hence, we can conclude that Eq. (\ref{exma1}) exhibits both conservative and dissipative like behaviour depending on the value of the parameter $k_3$. For $k_3>0$ Eq. (\ref{exma1}) exhibits isochronous nature; however for $k_3\leq 0$ we get dissipative nature.
\section{Comparison with other methods}

In  Ref.\cite{mus2}, Musielak has given the conditions to obtain the nonstandard Lagrangian of the form
\begin{equation}
L(x,\dot{x})=\frac{1}{H(x)\dot{x}+G(x)x}\label{lag1}
\end{equation}
for the equation of motion
\begin{equation}
\ddot{x}+f(x)\dot{x}^2+g(x)\dot{x}+h(x)x=0,\label{eq121}
\end{equation}
provided the following condition is satisfied
\begin{equation}
\frac{2}{3}g(x)=\frac{d}{dt}(3x\frac{h(x)}{g(x)})+3xf(x)\frac{h(x)}{g(x)},\label{eq2m2}
\end{equation}
following the earlier work on MEE equation \cite{vkc5}. The functions $H(x)$ and $G(x)$ are given by
\begin{equation}
H(x)=e^{I_f(x)}, G(x)=3\frac{h(x)}{g(x)}e^{I_f(x)}
\end{equation}
with
\begin{equation}
I_f(x)=\int f(x) dx.
\end{equation}
A similar approach is also given by Cie$\acute{s}$linski and  Nikiciuk in Ref.\cite{mus2new}.
We have applied the conditions given in Refs. \cite{mus2} and \cite{mus2new} to the present case of Eq.(\ref{int4}). This equation as well as the examples $1$ and $2$ (vide Eqs.(\ref{exm1}) and (\ref{exma1})) satisfy the above condition (\ref{eq2m2}) and yields the same form of Lagrangian (\ref{lag1}) (upto a constant multiple). However, in the present work, we have also brought out the interconnections in a systematic way and obtained the isochronous solutions.
\section{Conclusion}
In this paper, we have discussed the nonstandard Lagrangian and Hamiltonian structures of an isochronous nonlinear oscillator equation (\ref{int4}) belonging to the mixed type Li\'enard equation. We have utilized the interrelation between the Jacobi last multiplier method and Lie point symmetries and obtained the associated Lagrangian. However, the obtained Lagrangian is found to be too complicated. Hence, we utilized the interconnection between the PS method and last multiplier and obtained a time independent integral for Eq. (\ref{int4}) which in turn gives the nonstandard Lagrangian and Hamiltonian for Eq (\ref{int4}). Further, we have shown that the obtained Hamiltonian for Eq. (\ref{int4}) can be transformed to the Hamiltonian of a linear harmonic oscillator through appropriate canonical transformation from which we obtained the solution for Eq. (\ref{int4}). We have also shown that the system exhibits isochronous nature of oscillations for $\lambda>0$. We have also compared our results with the other existing methods and confirmed their correctness. The identification of such systems are of considerable interest from practical applications as many of these equations are of {\it PT}-symmetric type. Also, it is of considerable interest to consider the quantization of the system (\ref{int4}) and its generalization to higher dimensions. Hence, we believe that exploring such nonlinear equations will be highly rewarding in understanding general nonlinear systems.

\section{Acknowledgments}
AKT is very grateful to Centre for Nonlinear Dynamics, Bharathidasan University, Tiruchirappalli-620024 for warm hospitality. The work of VKC is supported by the SERB-DST Fast Track scheme for young scientists under Grant No. YSS/2014/000175. The work of MS forms part of a research project sponsored by Department of Science and Technology, Government of India. The work of ML is supported by a DAE Raja Ramanna Fellowship.


\end{document}